\documentclass[namedreferences]{solarphysics}
\usepackage[solaromanenum]{spr-sola-addons} % For Solar Physics
\usepackage{graphicx}                    % For eps figures, newer & more powerfull
\usepackage{color}                       % For color text: \color command
\usepackage[pdfborder={0 0 0 },urlcolor=blue,breaklinks]{hyperref}

\begin{document}
\begin{article}
\begin{opening}

\title{A Test of the Active Day Fraction Method of Sunspot Group Number Calibration: Dependence on the Level of Solar Activity}

%%%%%%%%%%%%%%%%%%%%%%%%%%%%%%%%%%%%%%%%%%%%%%%%%%%
%% Authors Names
%
\author{T.~\surname{Willamo}$^{1}$ \sep
    I.G.~\surname{Usoskin}$^{2,3}$ \sep
    G.A.~\surname{Kovaltsov} $^{4,2^\star}$
    }

%%%%%%%%%%%%%%%%%%%%%%%%%%%%%%%%%%%%%%%%%%%%%%%%%%%
%% Runningheads
%
\runningauthor{T. Willamo \textit{et al.}}
\runningtitle{Active Day Fraction: Dependence of Solar Activity}

%%%%%%%%%%%%%%%%%%%%%%%%%%%%%%%%%%%%%%%%%%%%%%%%%%%
%% Affilations
%
\institute{ $^1${Department of Physics, University of Helsinki, 00014 Helsinki, Finland}\\
$^2${Space Climate Research Unit, University of Oulu, Finland.}\\
 $^3${Sodankyl\"a Geophysical Observatory, University of Oulu, Finland.}\\
 $^4${Ioffe Physical-Technical Institute, St. Petersburg, Russia.}\\
 $^\star${(visiting scientist)}\\
                     email: {Ilya.Usoskin@oulu.fi}
             }

%%%%%%%%%%%%%%%%%%%%%%%%%%%%%%%%%%%%%%%%%%%%%%%%%%%
%%% Abstract
\begin{abstract}
The method of active day fraction (ADF) was proposed recently to calibrate different solar observers to
 the standard observational conditions.
The result of the calibration may depend on the overall level of solar activity during
 the observational period.
This dependency is studied quantitatively using data of the Royal Greenwich Observatory, by formally
 calibrating synthetic pseudo-observers to the full reference dataset.
It is shown that the sunspot group number is precisely estimated by the ADF method for periods of
 moderate activity, may be slightly underestimated by 0.5\,--\,1.5 groups ($\leq$10\,\%) for strong and
 very strong activity, and is strongly overestimated by up to 2.5 groups ($\leq$30\,\%) for weak--moderate activity.
The ADF method becomes unapplicable for the periods of grand minima of activity.
In general, the ADF method tends to overestimate the overall level of activity and to reduce
 the long-term trends.
\end{abstract}


\end{opening}

\section{Introduction}
\label{Sec:Intro}
Calibration of different solar observers to compile a long, homogeneous sunspot number series is
 of great importance now, when the ``classical'' sunspot-number series have been found partly erroneous
 \citep{clette14,usoskin_LR_17}.
Different methods of re-calibration of the sunspot number series were proposed
 \citep{clette14,lockwood_1_14,svalgaard16,clette16,usoskin_ADF_16,chatzistergos17}.
While most of the new re-calibrations were based on the direct pairwise or ``backbone'' comparison
 of individual observers, forming a daisy-chain of calibrations with error accumulation in time,
 an alternative active-day-fraction (ADF) method was proposed by \citet{usoskin_ADF_16} \citep[refined by][]{willamo17}.
The ADF method is based on a comparison of the statistic
 of active day (days when at least one sunspot group was observed, as opposed to quiet days with no spots) occurrence
 for each observer to that of the reference dataset, which is the Royal Greenwich Observatory (RGO) series for the period
 1900\,--\,1976 \citep[see details given by][]{usoskin_ADF_16}.
As such, the method is free of daisy-chaining and error accumulation since each observer is compared directly to the reference dataset.
As a drawback, however, the method may be somewhat sensitive to the overall level of solar activity during the period
 when the calibrated observer was making observations.
This was qualitatively mentioned in our earlier works \citep{usoskin_ADF_16,willamo17} but never investigated in
 sufficient detail.
This issue has been actively discussed during several in-person meetings and teleconferences of a group
 working on the sunspot-number series re-calibration, and it has also led to some unpublished critique \citep{svalgaard17}.

Accordingly, here we perform a full quantitative test of the dependence of the ADF method on the average
 solar-activity level during the period of observations of a calibrated observer.

\section{Method}

\subsection{RGO Pseudo-Observers}

The ADF method is tested using pseudo-observers based on the reference RGO dataset
 (available at solarscience.msfc.nasa.gov/greenwch.shtml).
From the entire RGO dataset we have selected several subsets, each covering three solar cycle max-to-max
 intervals, with start times separated by one solar cycle, as listed in Table~\ref{Tab:cycles}.
The periods covered by these pseudo-observers, correspond to different levels of solar activity,
 ranging from an average of 3.4 sunspot groups during the activity minimum around the turn of 19th and 20th
 centuries, to 6.15 groups during the Modern grand maximum.
Each of these subsets was considered as a pseudo-observer and calibrated to the full reference dataset
 following the procedure described by \citet{willamo17}.
This approach makes it possible to directly assess possible biases of the method.
Since the pseudo-observers are simply subsets of the reference dataset, the formal calibration should be
 (consistent with) zero: any non-zero difference between the ``calibrated'' and the reference
 data indicates a bias in the calibration procedure.

\subsection{Observational Thresholds}

In the framework of the ADF method, each observer is characterized by their observational threshold ($S$),
 which is the minimum observed (\textit{i.e.} uncorrected for foreshortening) area (in millionths of the solar disc: msd)
 of a sunspot group resolvable by the observer.
In other words the observer is supposed to record all groups larger than, and miss all groups smaller than, $S$.
The value of $S$, along with its uncertainties, is defined by matching the statistic (cumulative probability
 distribution function; cpdf) of the ADF occurrence recorded by the observer with that of the reference dataset
 by applying to the latter the observational threshold \citep[see full details in][]{willamo17}.
This assumes that the calibrated observer is ``poorer'' than the reference observer, meaning that the positive
 threshold is applied to the reference dataset.

The formal application to the ADF calibration procedure to the pseudo-observers 1, 2 and 3 (see Table~\ref{Tab:cycles})
 yields the thresholds of 15 to 45 msd, implying that these pseudo-observers appear poorer than the reference set.
This can be understood as follows.
Of course, all pseudo-observers must have the same quality (\textit{i.e.} $S=0$) by construction, as subsets of the
 reference dataset.
However, because of the low level of solar activity, there are fewer active days (lower ADF) during time interval 1,
 which is interpreted by the ADF method as a lower quality of the observer, or in other words, a positive threshold $S$.
This formal threshold is a bias of the method, related to the low level of activity, which is below the mean
 group number $\langle G\rangle=4.27$ for the reference dataset.
It appears large for pseudo-observers 1 and 2 but small, almost consistent with no threshold, for pseudo-observer 3.

The situation is inverted for pseudo-observers 4 through 6, which correspond to time intervals with enhanced
 solar activity $\langle G\rangle>4.27$.
Because of that, there are more active days during these periods, which is interpreted by the method as
 a quality of the pseudo-observers higher than that of the reference dataset.
This would lead to formally negative thresholds, which however does not make sense, since the sunspot group area
 cannot be negative.
In order to deal with this, we define the negative threshold as follows.
The threshold $S$ is considered negative when it is applied to the calibrated observer data, while keeping the reference
 dataset as it is (no threshold).

\begin{table}
\caption{Three-cycle pseudo-observers used in this study for ``reconstruction'' of RGO-based group number ($G$).
 Columns present the time intervals, the mean value of $G$ during the interval, the obtained observational
 threshold ($S$) with 68\,\% confidence interval. }
%\begin{center}
\begin{tabular}{cccc}
\hline
Pseudo-observer & Time [years] & $\langle G\rangle $& $S$ [msd]\\
\hline
1&1883.9 -- 1917.6 & 3.43 & $46\left(_{41}^{52}\right)$\\
2&1894.1 -- 1928.4 & 3.65 & $30\left(_{24}^{34}\right)$\\
3&1907.0 -- 1937.4 & 4.01 & $14\left(_{10}^{17}\right)$\\
4&1917.6 -- 1947.5 & 4.65 & $-15\left(_{-18}^{-11}\right)$\\
5&1928.4 -- 1957.9 & 5.4 & $-14\left(_{-17}^{-10}\right)$\\
6&1937.4 -- 1968.9 & 6.15 & $-38\left(_{-44}^{-29}\right)$\\
\hline
\end{tabular}
%\end{center}
\label{Tab:cycles}
\end{table}

\begin{figure}
\centering
\includegraphics[width=1.0\columnwidth]{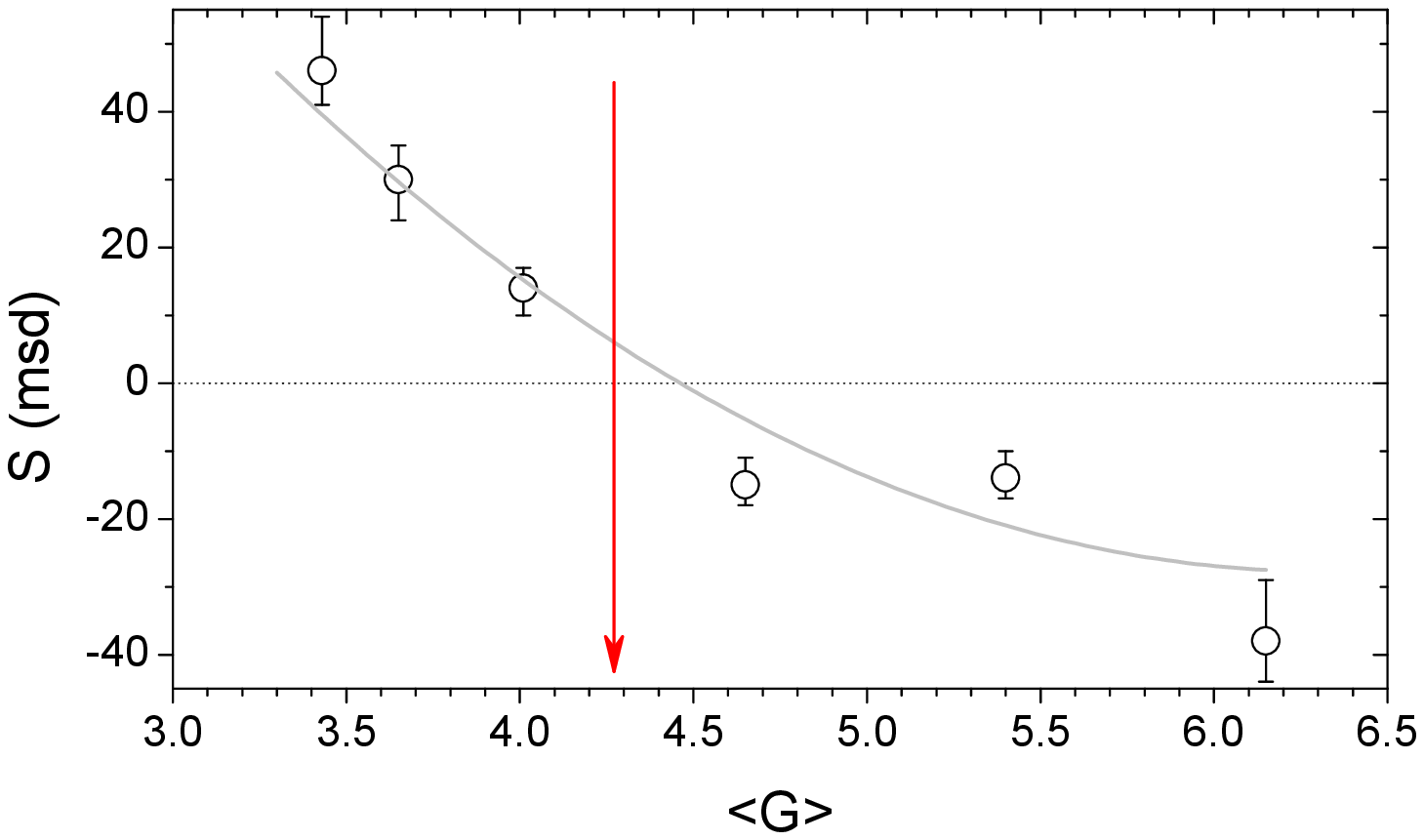}
\caption{The observational threshold $S$ derived for the pseudo-observers (see text) for the
 six intervals given in Table~\ref{Tab:cycles} with the 68\,\% confidence intervals.
 An eye-guiding best-fit parabolic curve is given in grey to emphasize the non-linearity of the relation.
 The red arrow denotes the mean $\langle G\rangle=4.27$ value for the period 1900\,--\,1976.}
\label{fig:S}
\end{figure}
Thus-defined values of $S$ are given in Table~\ref{Tab:cycles} and presented in Figure~\ref{fig:S}.
One can see that the values of $S$ are positive for low-activity cycles (points left of the red arrow in the Figure)
 and negative for high-activity cycles, as expected.
An interesting feature is that the relation between $S$ and $\langle G\rangle$ is strongly non-linear, suggesting
 that the method is more accurate for moderate and high cycles than for low cycles.
An eye-guiding line (best weighted-fit parabola) is shown for illustration.

\subsection{Data ``Correction''}

\begin{figure}
\centering
\includegraphics[width=1.0\columnwidth]{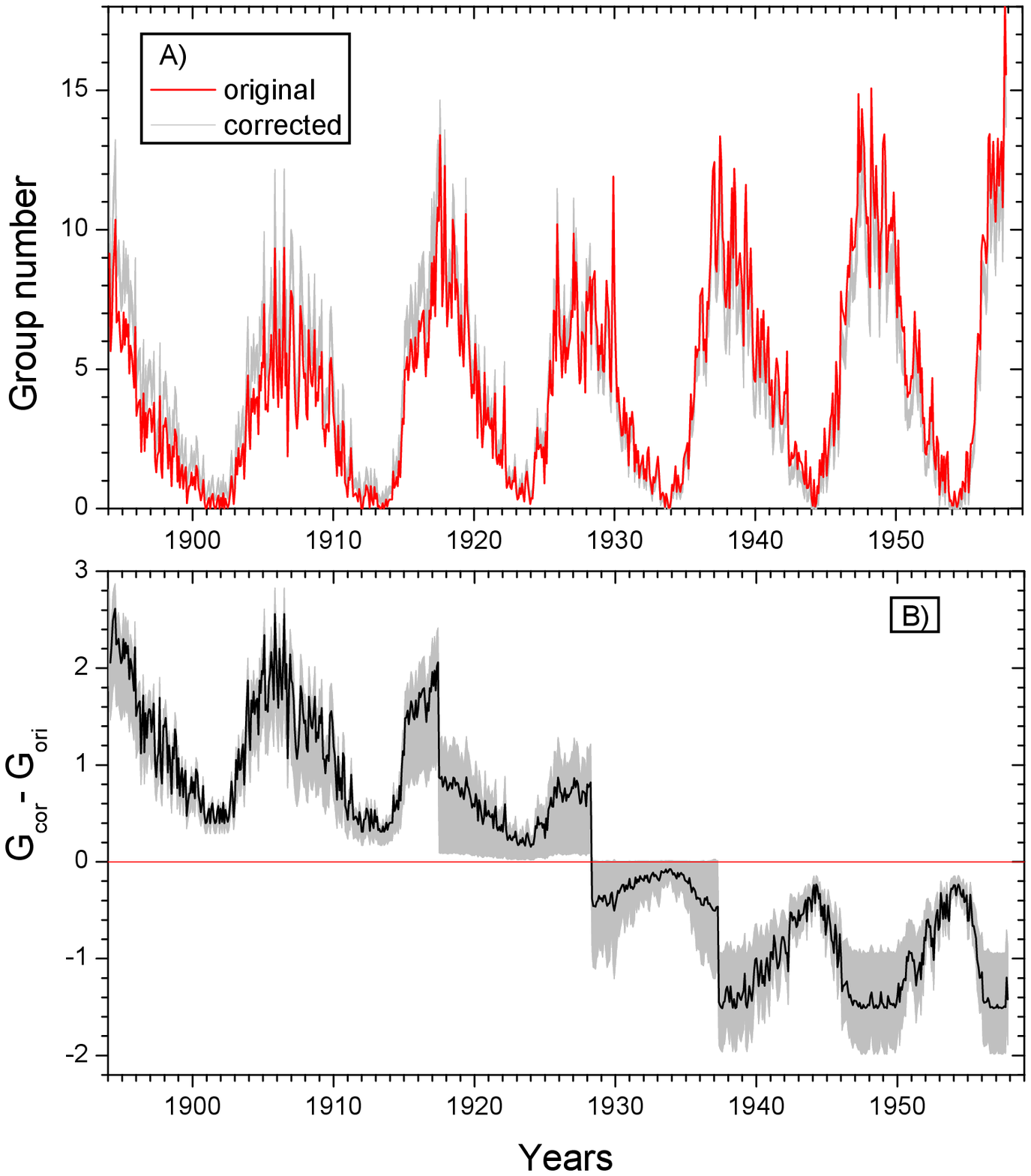}
\caption{Comparison between original [$G_{\rm ori}$, red] and ``corrected'' [$G_{\rm cor}$, grey, see text] monthly group numbers
 for the RGO dataset.
Panel A: temporal variability of the two series.
Panel B: The difference between the two series.
Gray shading denotes the 68\,\% ($1\sigma$) confidence intervals.}
\label{fig:Diff}
\end{figure}

Next we ``corrected'' the data of pseudo-observers, using the values of the threshold obtained above (Table~\ref{Tab:cycles})
 and applying the procedure described earlier \citep{usoskin_ADF_16,willamo17}.
For negative values of $S$, the correction corresponding to $-S$ was applied by inverting its sign (not added to but subtracted
 from the reported $G-$values).
Corrections were applied to the RGO data for the corresponding pseudo-observers from Table~\ref{Tab:cycles}.
Since they overlap in time, we apply the average over them for each day, as described by \citet{willamo17}.
The resultant ``corrected'' series is shown (with the 68\,\% confidence interval) in Figure~\ref{fig:Diff}A in grey.
One can see that the ``corrected'' series is slightly higher than the original one in the early part
 and slightly lower in the later part of the plot.
The difference between the ``corrected'' and the original series is shown in panel B.
Jumps are caused by the boundaries between pseudo-observers.

The group number ($G$) is significantly overestimated for low solar cycles (1894\,--\,1917), with the
 difference reaching 2.5 sunspot groups.
Corrections are small and consistent with zero within the 68\,\% confidence interval for moderate cycles
 (1917\,--\,1937).
Group number appears slightly underestimated, up to 1.5 groups, for high cycles (1937\,--\,1957), which include
 also the highest known solar cycle: No. 19.

\section{Discussion and Conclusions}

We have analyzed dependence of the ADF method of sunspot-group number re-calibration on the level of solar activity
 during the observational period of an observer.
Such a dependence was discussed earlier \citep{usoskin_ADF_16} but not assessed quantitatively.
Using fragments of the reference dataset of RGO we formed several pseudo-observers, who were formally calibrated
 against the full reference dataset.
Since the pseudo-observers, by construction, are identical to the reference dataset, the true calibration must be zero
 or consistent with zero within uncertainties.
Accordingly, any non-zero formal ``correction'' of the pseudo-observers can be interpreted as a bias in the method.
We found that indeed, as proposed earlier, the ADF method tends to overestimate weak solar activity and underestimate high activity.
However, the bias introduced is non-linear.
While moderate cycles with the peak values of $G=10$\,--\,12 are reproduced correctly, high and very high cycles $G=$12\,--\,17
 may appear underestimated by 0.5\,--\,1.5 groups, or less than 10\,\%.
On the other, hand, weak-moderate cycles with $G$=7\,--\,9 appear overestimated by 2\,--\,2.5 groups, or about 30\,\%.
As discussed by \citet{usoskin_ADF_16}, the ADF method cannot work with a very weak cycle, e.g. during the Dalton minimum.
It is important to notice that this is related not to the height of individual cycles but to the average level of solar
 activity during the entire time span of a given observer.

As a result, in the long run, the ADF method tends to overestimate the overall level of activity and to reduce
 the long-term trends.
Therefore, the sunspot-activity level reconstructed by the ADF method can be considered as a conservative upper
 limit for estimates of the long-term trends of solar activity.

In summary we conclude that:
\begin{itemize}
\item
The ADF method works accurately for the period of moderate solar activity.
\item
Application of the ADF method may lead to a slight underestimate, by 0.5\,--\,1.5 sunspot groups ($\leq$10\,\%), of
 solar activity during periods of high and very high activity.
\item
Application of the ADF method may lead to a significant overestimate,  up to 2.5 sunspot groups ($\leq$30\,\%), of
 solar activity during periods of low-moderate activity.
\item
Overall, the ADF method tends to overestimate the overall level of activity and reduce
 the long-term trends.
\end{itemize}

%==========================
\section*{Acknowledgements}
We are thankful for fruitful discussion to Fr\'ed\'eric Clette, Ed Cliver, Greg Kopp, Laure Lef\'evre, Andr\'es M\~unoz-Jaramillo,
Alexei Pevtsov, Leif Svalggard, and Jos\'e Vaquero.
VarSITI/SCOSTEP is acknowledged for an opportunity to discuss this material during the 2nd VarSITI General Symposium (Irkutsk, 2017)
This work was carried out in the framework of ReSoLVE Centre of Excellence (Academy of Finland, project no. 272157).

\section*{Disclosure of Potential Conflicts of Interest}
The authors declare that they have no conflicts of interest.

%===================================================
%\bibliographystyle{spr-mp-sola} % style aa.bst
%\bibliography{J:/usoskin/papers/usoskin_all}
%\bibliography{C:/DATA__/USOSKIN/papers/usoskin_all}
%\bibliography{usoskin_all}
%\bibliographystyle{plainnat}

\end{article}
\end{document}